\documentclass[twocolumn,pra]{revtex4}

\usepackage{graphicx}% Include figure files
\usepackage{dcolumn}% Align table columns on decimal point
\usepackage{bm}% bold math

\newcommand{\beq}{\begin{equation}}
\newcommand{\eeq}{\end{equation}}
\newcommand{\beqa}{\begin{eqnarray}}
\newcommand{\eeqa}{\end{eqnarray}}
\newcommand{\beqar}{\begin{eqnarray*}}
\newcommand{\eeqar}{\end{eqnarray*}}

\begin{document}
\input epsf
\title{\bf \large Entangling and disentangling capacities of nonlocal maps}

\author{Berry Groisman}
\affiliation{Centre for Quantum Computation, DAMTP, Centre for
Mathematical Sciences, University of Cambridge, Wilberforce Road,
Cambridge CB3 0WA, United Kingdom. }

\begin{abstract}
Entangling and disentangling capacities are the key manifestation of
the nonlocal content of a quantum operation. A lot of effort has
been put recently into investigating (dis)entangling capacities of
unitary operations, but very little is known about capacities of
non-unitary operations. Here we investigate (dis)entangling
capacities of unital CPTP maps acting on two qubits.
\end{abstract}

\maketitle

%\pacs{03.67-a, 03.65.Ud, 03.65.Ta}

\section{Introduction}\label{intro}
Entanglement content is one of the fundamental ways to characterize
nonlocal quantum resources (nonlocal states and operations). For
pure bipartite states the ultimate measure of entanglement, the von
Neumann entropy of entanglement, had been recently discovered
\cite{proc_meth}. A universal measure of entanglement for mixed
states had not been found yet and different measures are used
depending on the operational context. Nevertheless, the important
feature of all entanglement measures of states is that their values
are directly inferred using the parameters of a state itself.

Similarly to mixed states, the entanglement content of quantum
operations can be characterized in different ways, e.g. via the
amount of entanglement necessary to generate that operation or via
the amount of entanglement the operation is able to produce/destroy
(the so called entangling/disentangling capacities). This article is
concerned with the two latter measures.

Unlike the amount of entanglement in a state, the (dis)entangling
capacities of an operation do not have an operational interpretation
on their own. They manifest themselves via the change of the
entanglement of a particular state that the operation acts upon. And
the operation has to act on a specific state (the ``optimal" state)
in order to realize its (dis)entangling capacity in full. Thus, the
straightforward way to calculate these quantities is to maximize the
change of entanglement over all possible initial states.

Substantial progress have been made recently in investigating
(dis)entangling capacities of unitary operations. The capacities of
two-qubit unitary operations were explicitly calculated \cite{nlham,
lhl, BS_2005}. It was also shown that single-shot capacities are
equal to asymptotic capacities \cite{lhl,BHLS_cap_Ham,asym_ent_cap}.
Some results for higher dimensions were also obtained
\cite{ent_dis_unequal}. However, extending these techniques to
systems of higher dimensionality seems to be a very difficult task.
Even in the two-qubit case the capacities of a general unitary
operation have been calculated numerically, no analytical technique
is known.

In all real situations an experimentalist never deals with a perfect
unitary in the laboratory. And it is needless to say that
calculating capacities of non-unitary operations, i.e. nonlocal
quantum maps, is even a bigger challenge.

In this article we consider nonlocal completely positive trace
preserving (CPTP) maps of the form

\begin{equation}\label{map}
    \tau(\rho)\rightarrow\sum_k p_k U_k \rho U_k^\dag,
\end{equation}
where $U_k$ are unitary transformations. Maps of this type are often
called {\it random unitary processes}, and they are doubly
stochastic. The map (\ref{map}) may arise, for example, if the
desired unitary transformation can be implemented successfully only
with certain probability, while another unitary is realized in the
case of failure. A continuous version of the map (\ref{map}) may
arise naturally in experiment if parameters of a desired unitary
transformation are subject to a noise (the case of Gaussian noise
will be analyzed in detail in Section \ref{sec:continuous}). The
scope of this article covers the case of $\tau$ that act on two
qubits. We calculate single-shot (dis)entangling capacities of
$\tau$ in some particular cases.

 The structure of the article is as follows. In Sec.
 \ref{sec:unitary} the definition(s) of (dis)entangling
 capacities of unitaries are presented and some recent results concerning two-qubit
 unitaries.
 Sec. \ref{sec:nonunitary} generalizes the definition of (dis)entangling
 capacities for non-unitaries.
Some numerical results for (lower bounds on) (dis)entangling
capacities for discreet and continuous mixtures of unitaries are
presented in Sec. \ref{sec:examples}.

\section{(Dis)entangling capacity of a unitary:
Definitions and some related results}\label{sec:unitary} Consider an
unitary operation $U_{AB}$ that acts on a tensor product Hilbert
space $\mathcal{H}_A\otimes\mathcal{H}_B$ of two spatially separated
particles $A$ and $B$. If $U_{AB}$ can not be decomposed into a
tensor product of local unitaries, i.e. $U_{AB}\neq V_A \otimes
W_B$, then we say that $U_{AB}$ is nonlocal. Unlike local unitaries,
nonlocal unitaries have an ability to produce or destroy
entanglement. This ability is usually characterized by the
entangling, $E^\uparrow(U)$, and the disentangling,
$E^\downarrow(U)$, capacities, i.e. by the maximal increase
(decrease) of entanglement that can be achieved when $U$ acts on
quantum states. To quantify these capacities we have to choose
appropriate measures of entanglement. The most sensible choice is to
use {\it the entanglement of formation} \cite{wootters} as a measure
of entanglement of the initial state $\rho$, and {\it the
distillable entanglement} \cite{dist_ent:BDiVSW} as a measure of
entanglement of the final state $U\rho U^\dagger$. The reason for
this asymmetric choice is purely operational one. What counts is the
amount of resources (pure maximally entangled states) needed to
create $\rho$ (asymptotically) and the amount of pure-state
entanglement one will be able to extract from $U\rho U^\dagger$,
again asymptotically. Thus the most general definition is

\parbox{0.4in}{\begin{eqnarray*}\end{eqnarray*}}\hfill
\parbox{2.2in}{\begin{eqnarray*} \label{def_U_mixed}
E^\uparrow(U)=\max_{\rho} [D(U\rho U^\dagger)-E_F(\rho)],\\
E^\downarrow(U)=\max_{\rho} [E_F(\rho)-D(U\rho U^\dagger)],
\end{eqnarray*}}\hfill
\parbox{0.5in}{\begin{eqnarray}\end{eqnarray}}

\noindent where the maximization is over all possible states $\rho$
(mixed and pure) accessible to $U$. The Hilbert space of an
accessible $\rho$ is not necessarily restricted to
$\mathcal{H}_A\otimes\mathcal{H}_B$. It turns out to be the case
that some $U$ create more entanglement if the original particles are
entangled with local ancillary particles \cite{nlham,lhl}.
%are defined as the maximal increase (decrease) of entanglement when
%$U$ acts on a {\it pure} state \cite{pure}
It also appears to be the case that the maximization in Eq.
(\ref{def_U_mixed}) can be restricted to pure-states only
\cite{lhl},
%In reaching this conclusion
%\cite{lhl} used the fact that $E_F$ is a convex measure.
therefore, the definition (\ref{def_U_mixed}) can be simplified as

\parbox{0.4in}{\begin{eqnarray*}\end{eqnarray*}}\hfill
\parbox{2.2in}{\begin{eqnarray*}\label{def_U_pure}
E^\uparrow(U)=\max_{\psi} [E(U|\psi\rangle)-E(|\psi\rangle)]~\nonumber\\
E^\downarrow(U)=\max_{\psi} [E(|\psi\rangle)-E(U|\psi\rangle)],
\end{eqnarray*}}\hfill
\parbox{0.5in}{\begin{eqnarray}\end{eqnarray}}

\noindent where $E$ is an entanglement measure for pure state
(Throughout this paper we will use the von Neumann entropy of
entanglement as the most appropriate measure). This obviously
simplifies the job significantly.

Let us briefly recall the main results for $A$ and $B$ being
two-level particles, qubits.

Any $U_{AB}$ acting on qubits can be decomposed as
\cite{nlham,krauscirac}
\begin{equation}
U_{AB}=\left[V_A\otimes V_B \right ]e^{i\sum_{\alpha=x,y,z}
\xi_{\alpha} \sigma_\alpha^A\sigma_\alpha^B} \left [W_A\otimes
W_B\right ],
\end{equation}
where $\pi/4\geq\xi_x\geq\xi_y\geq|\xi_z|\geq0$. The middle term
sandwiched by local unitaries is called the {\it canonical
decomposition} of $U$. Any $U$ can be transformed to its canonical
form by sandwiching it with Hermitian conjugates of corresponding
local unitaries. That means that the canonical form is genuinely
nonlocal part of $U$ - everything else is local. The beauty of this
results is that out of 15 real parameters that parameterize a
general two-qubit unitary only three are necessary to describe its
nonlocal nature. It simplifies considerably the classification of
nonlocal unitaries. For the purpose of our discussion three classes
can be identified; namely, the Controlled-NOT(CNOT)-class
($\xi_x\neq 0$, $\xi_y=\xi_z =0$), the DoubleCNOT-class ($\xi_x\neq
0$, $\xi_y\neq 0$, $\xi_z=0$), and the SWAP-class (all three
$\xi_{\alpha}$ are not equal zero) \cite{classification,U->U}. The
names reflect the fact that the corresponding ``mother" unitary
transformation (i.e. with $\xi_{\alpha}=\pi/4$ for $\alpha\neq0$)
belongs to that class.

The main results for qubits are \cite{lhl,nlham}:

(a) $E^\uparrow(U)=E^\downarrow(U)$.

(b) For CNOT-class the optimal state, i.e. the state that satisfies
definition (\ref{def_U_pure}), lives solely in the Hilbert space of
particles $A$ and $B$ (no ancillas are needed) and takes the form
\begin{equation}\label{input}
|\psi^{opt}\rangle=\cos\alpha|0\rangle_A|0\rangle_B\pm i \sin\alpha
|1\rangle_A|1\rangle_B,
\end{equation}
where $\pm$ correspond to $E^\uparrow$ and $E^\downarrow$
respectively. Thus all $U$ from that class achieve their capacity by
acting on pure states with the same Schmidt basis (only values of
Schmidt coefficients differ depending on the value of $\xi_x$).
 The values $\alpha=f (\xi_x)$ can be obtained by
straightforward numerical optimization.

(c) If $\xi_{\alpha}<\pi/4$ the maximal capacity is achieved when
$|\psi^{opt}\rangle$ is already entangled.

(d) Unitaries of the CNOT-class achieve their capacities by acting
on optimal states that lie in $\mathcal{H}_A\otimes\mathcal{H}_B$.
However, unitaries of the DCNOT and SWAP-classes achieve their
capacities only if the original particles are entangled with local
ancillas. It was conjectured that it is sufficient to take the size
of ancillas equal to the size of original particles. This conjecture
was supported by numerical simulations for qubits \cite{lhl}.

%On the one hand, it might seem not surprising at all. Indeed, in the
%definition (\ref{def_U_mixed}) the bipartite Hilbert space in which
%$\rho$ lives is not restricted. Let us assume that the maximum is
%achieved for some state $\rho_{\ast}$. However, $\rho_{\ast}$ can
%always be purified $|\psi_{\ast}\rangle$ by increasing a local
%Hilbert space. In other words, for any mixed state $\rho$ there
%exist a corresponding pure state, that achieves the same increase of
%entanglement.

%I define $\overline{E_c(\tau)}$ and $\overline{D_c(\tau)}$ as the
%maximal average increase (decrease) of entanglement when $\tau$ acts
%on {\it pure} state:
%\begin{eqnarray}
%\overline{E_c(\tau)}=\max_{\psi} \sum_i p_i
%[E(U_i|\psi\rangle)-E(|\psi\rangle)]\nonumber\\
%\overline{D_c(\tau)}=\max_{\psi} \sum_i p_i
%[E(|\psi\rangle)-E(U_i|\psi\rangle)]
%\end{eqnarray}

\section{Entangling and disentangling capacities of a
non-unitary}\label{sec:nonunitary}

For non-unitaries we will use a definition similar to Eq.
(\ref{def_U_mixed}), i.e. we define

\parbox{0.1in}{\begin{eqnarray*}\end{eqnarray*}}\hfill
\parbox{2.6in}{\begin{eqnarray*}\label{def_tau_mixed}
E^\uparrow(\tau)=\max_{\rho} [D(\tau(\rho))-E_F(\rho)]~\nonumber\\
E^\downarrow(\tau)=\max_{\rho} [E_F(\rho)-D(\tau(\rho))].
\end{eqnarray*}}\hfill
\parbox{0.5in}{\begin{eqnarray}\end{eqnarray}}

However, in general here we cannot justify reducing the search to
pure states. This is due to the fact that distillable entanglement
is not necessary a convex measure.

We can argue, nevertheless, that in the case of mixtures of
unitaries acting on two qubits without ancillas the distillable
entanglement can be regarded as a convex measure. Indeed, a mixture
of optimal states (\ref{input}) forms a Bell-diagonal state for
which the lower and upper bounds on distillable entanglement
\cite{VP:ent_m_pp}
\begin{equation}
S(\rho_A)-S(\rho_{AB})\leq D(\rho)\leq E_{RE}(\rho_{AB})
\end{equation}
 coincide. Here we recall that the relative entropy of entanglement $E_{RE}(x)$
 is a convex measure.

If ancillas are used then the situation is more complicated. We
leave the question of whether the capacities are attained on pure
states as an open question and calculate the lower bounds on these
capacities using pure states.

\section{Mixtures of unitaries acting on two qubits}\label{sec:examples}

The properties of two-qubit unitaries described above in Section
\ref{sec:unitary} can help us to generalize that approach to
mixtures of unitaries as in Eq.
(\ref{map})\cite{mixture_same_class}. Here we use two methods for
calculating $E^{\uparrow}(\tau)$ and $E^{\downarrow}(\tau)$.

{\bf Method I:} We make an assumption about the particular form of
the optimal input state, and subsequently find the optimal values of
its parameters.

{\bf Method II:} We perform a direct numerical optimization without
making any {\it a priori} assumption about the optimal state (except
of its purity).

\subsection{Example I: Discreet CNOT-mixtures}\label{sec:example_I}
Consider a mixture of unitaries of the CNOT-class,
\begin{equation}\label{cnot}
U_k=e^{i \xi_x^k  \sigma_x^A\sigma_x^B}.
\end{equation}
Here we use Method I. From continuity it follows that the optimal
state is expected to lie on the 2-dimensional manifold of
(superpositions of) states of the form (\ref{input}) or their convex
mixtures. Moreover, in this special case we can adopt the argument
of \cite{lhl} (see Sec. \ref{sec:unitary}) and claim that the search
can be restricted to pure states only. Thus, the state optimal for
$\tau$ is again of the type (\ref{input}).

As a simplest case let us consider only two unitaries $U_1$ and
$U_2$:
\begin{equation}\label{map2}
%---source: entanglement capability.nb------
    \tau(\rho)\rightarrow p U_1 \rho U_1^\dag  + (1-p) U_2 \rho U_2^\dag
\end{equation}
\begin{figure} \epsfxsize=3.7truein
      \centerline{\epsffile{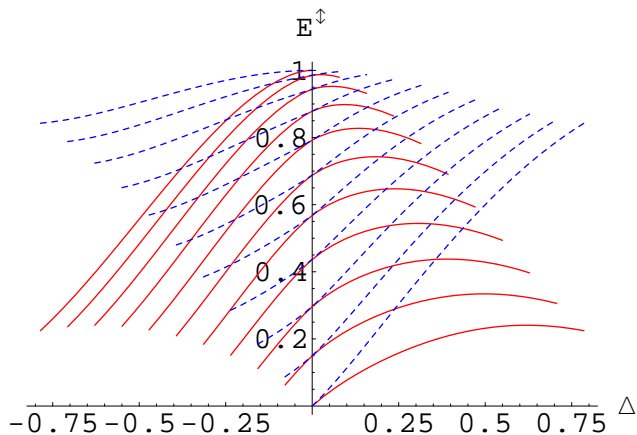}}
  \caption[]{(color online) $E^{\uparrow}(\tau)$ (solid line) and $E^{\downarrow}(\tau)$ (dashed line) as functions
   of $\Delta$ for different values of $\xi$, where $p=0.5$.
  The highest curve corresponds to $\xi=\pi/4$. The lowest curve corresponds to $\xi=0$.
  Here $\Delta$ is measured in radians.}
    \label{fig1} \end{figure}
For convenience let us define $\Delta=\xi_x^2-\xi_x^1$, and denote
$\xi_x^1$ simply by $\xi$, so $\xi_x^2=\xi+\Delta$. We will fix
$\xi$ and analyze $E^\uparrow$ and $E^\downarrow$ for various
$\Delta$. For $\Delta=0$ the map reduces to a unitary (with an
appropriate capacity). As smaller angle means smaller
$E^\uparrow(U)$, we would expect that if $U_1$ is mixed with $U_2$,
where $\Delta<0$, then the entangling capacity of the resulting map
will decrease relative to $E^\uparrow(U)$. This intuition is
consistent with the results presented on Fig. \ref{fig1}. Similarly,
we might expect that the entangling capacity of the map will
increase with $\Delta>0$, and that this capacity will reach its
maximum for maximal $\Delta$, i.e. maximal $E^\uparrow(U_2)$.
However, Fig. \ref{fig1} shows that this is not the case.
$E^\uparrow(\tau)$ indeed grows while $\Delta$ is positive and
relatively small, reaching its maximum for certain intermediate
positive value of $\Delta$ and then starting to decrease. In other
words, if $U_1$ and $p$ are fixed, then maximal $E^{\uparrow}(\tau)$
is achieved for some intermediate $U_2$ with $\xi_x^2>\xi_x^1$, but
not for $\xi_x^2=\pi/4$. This result might seem counterintuitive
from the first sight, but it has a clear explanation. Let
$\alpha_1$, $\alpha_2$ be the corresponding optimal values of
$\alpha$ in Eq. (\ref{input}) for $U_1$ and $U_2$ respectively, then
the optimal value of $\alpha_{\tau}$ will lie somewhere in between,
i.e. satisfy $\alpha_1>\alpha_{\tau}>\alpha_2$. For $\xi^2_x=\pi/4$,
$U_2$ can realize its entangling capacity of 1 ebit if $\alpha_2=0$.
However, when $U_2$ acts on a state with $\alpha_{\tau}>\alpha_2$,
then it creates less than $1-H[(\cos\alpha_{\tau})^2]$ ebit.

Disentangling capacity, $E^{\downarrow}(\tau)$, behaves differently.
It is monotonic with $\Delta$. It equals $E^{\uparrow}$ for
$\Delta=0$ as expected, and it is strictly larger than
$E^{\uparrow}$ for all other values of $\Delta$. The last
observation shows behavior completely opposite to that of unitaries.
In a sense, it is easier for non-local map to destroy entanglement
rather than create it, while for unitary operations the opposite
holds \cite{ent_dis_unequal}.

This approach can be similarly applied to any finite number of
unitaries and to continuous distribution of unitaries, which will be
discussed in Section \ref{sec:continuous}.

\subsection{Example II: Discreet DCNOT and SWAP-mixtures}\label{example_II}
For mixtures of unitaries of DCNOT and SWAP-class Method II was
used. The details of numerical calculations are presented in
Appendix.
\begin{figure} \epsfxsize=3.5truein
%----source: search4x4states4entcapacity.nb----
      \centerline{\epsffile{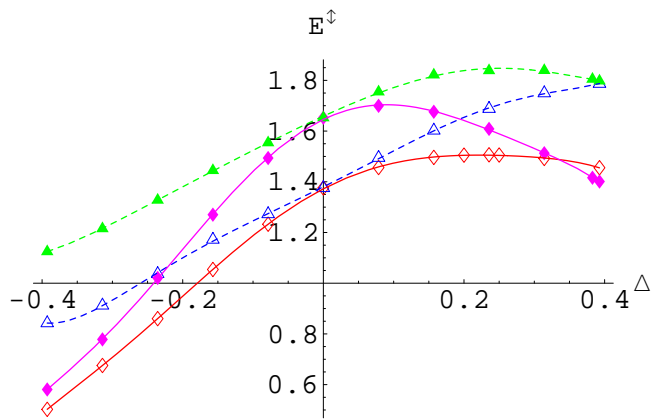}}
  \caption[]{(color online) DCNOT: $E^{\uparrow}(\tau)$ (empty diamonds with solid line fit) and $E^{\downarrow}(\tau)$
  (empty triangles with dashed line fit) as functions
   of $\Delta$ for $\xi_x^1=\xi_y^1=\pi/8$. SWAP: $E^{\uparrow}(\tau)$ (filled diamonds with solid line
fit)
   and $E^{\downarrow}(\tau)$
  (filled triangles with dashed line fit) as functions
   of $\Delta$ for $\xi_x^1=\xi_y^1=\xi_z^1=\pi/8$. In both cases $p=0.5$. Here $\Delta$ is measured in radians.}
    \label{fig2} \end{figure}
We conjectured that it is sufficient if local ancillas are qubits.
Selected results are shown in Fig. \ref{fig2}. We can see that for
DCNOT-mixture the behavior of $E^{\uparrow}(\tau)$ and
$E^{\downarrow}(\tau)$ is qualitatively similar to CNOT-mixture
(Fig. \ref{fig1}). However, for SWAP-mixture slightly different
behavior is obtained. In particular $E^{\downarrow}(\tau)$ exhibits
maximum at an intermediate value of $\Delta$. It is also noticeable
that for $D>0.357$, $E^{\uparrow}(\tau)$ is smaller for SWAP-mixture
than for DCNOT-mixture. It is a counterintuitive result that
SWAP-mixture which is naturally considered as ``stronger" that
DCNOT-mixture has lower entangling capacity. However, again similar
to the line of thought in Example I we can argue that for
(relatively) large $\Delta$ the second unitary, $U_2$ is too strong,
and therefore when it acts on the optimal state (optimal for the
mixture, not for itself) it causes more destruction that
corresponding $U_2$ of DCNOT-class would have caused.

\subsection{Example III: Entangling capacity of noisy unitary with Gaussian
fluctuations}\label{sec:continuous} So far we analyzed discrete
mixtures of unitaries. In this section we analyze a continuous
distribution, which is usually what experimentalists deal with.
These distributions arise due to uncertainty in one or more
parameters. Such uncertainties may be caused by the limits of
calibration precision of the devices and by high sensitivity of
systems used to generate desired interactions. For example, the
strength of exchange coupling between donors in silicon based
solid-state architectures for quantum computing exhibit significant
uncertainty resulting in error in gate operation \cite{silicon_CNOT:
THWH}.

 In particular, we
consider the case when a non-unitary map arises if a unitary from
CNOT-class is subject to a Gaussian noise.

 Recent work \cite{noisyHam} analyzed
the capability of noisy Hamiltonians to create entanglement. In
particular, interactions of the form Eq. (\ref{cnot}), where $\xi$
is Gaussian distributed with the mean $\overline{\xi}=\pi/4$ and
standard deviation $\Omega$, were considered. Without noise this
operation (CNOT operation) is able to create a maximally entangled
state if it acts on a disentangled pure state. The authors analyzed
the situation when the noisy operation acts on initially
disentangled state, which is by itself subject to a Gaussian noise.
Its capability to create entanglement was characterized in terms of
the condition for inseparability of the resulting mixed state (via
Peres-Horodecci separability criterion).

The aim of our analysis is different. We consider noisy interactions
with $\overline{\xi}\in [0,\pi/4]$ and calculate their entangling
and disentangling capacities in terms of $\overline{\xi}$ and
$\Omega$. Thus, we give a comprehensive quantitative
characterization of the non-local content of these noisy maps in
terms of their entangling and disentangling capacities. Unlike
\cite{noisyHam} we do not test the resulting state on
inseparability, rather calculate its distillable entanglement
explicitly.

\begin{figure} \epsfxsize=3.4truein
%--origin search$x4states4entcapacity.nb---
      \centerline{\epsffile{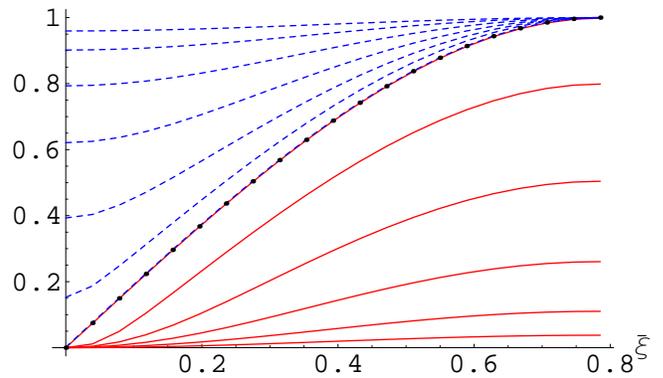}}
  \caption[]{(color online) $E^\uparrow(\tau_G)$ (solid line) and $E^\downarrow(\tau_G)$ (dashed line) as a function of $\overline{\xi}$
  for several values of $\Omega$: 0.01, 0.18, 0.35, 0.52, 0.69, and 0.86. The dotted line corresponds to $\Omega=0$, i.e. a
  unitary transformation. $\overline{\xi}$ and $\Omega$ are measured in radians.}
    \label{fig3} \end{figure}

The action of a unitary $U=\exp[i \xi \sigma_x^A \sigma_x^B]$, where
$\xi$ is Gaussian distributed with the mean $\overline{\xi}$ and the
standard deviation $\Omega$, on the state $\rho_{AB}$ can be seen as
a non-unitary CPTP map
\begin{equation}\label{Gmap}
    \tau_G(\rho)\rightarrow \frac{1}{\sqrt{2\pi}\Omega}
    \int_{-\infty}^{+\infty}e^{-\frac{(\xi-\overline{\xi})^2}{2\Omega^2}}U\rho
    U^\dag d\xi,
\end{equation}
which is a continuous mixture of unitaries of the CNOT-class.
Similarly to the Sec. \ref{sec:example_I} we consider pure initial
state, i.e. $\rho=|\psi\rangle\!\langle\psi|$, where $\psi$ takes
the form (\ref{input}), calculate the distillable entanglement of
the output mixed state, and maximize it over $\alpha$. Figure
\ref{fig3} presents numerical results for $E^{\uparrow}(\tau)$ and
$E^{\downarrow}(\tau)$ as functions of $\overline{\xi}$ for several
values of $\Omega$. We can see that already for
$\Omega\approx0.01~rad$ we obtain only very small deviation from the
(dis)entangling capacity of the unitary, i.e. $\tau_G$ without noise
- $\Omega=0$. As $\Omega$ increases the disentangling capacity
increases and the entangling capacity decreases.
The former fact should not be surprizing as it is known that
entanglement can be destroyed even by local CPTP unital maps
\cite{GPW}. Thus, the more dispersed the distribution of $\xi$
becomes, the easier for $\tau_G$ to destroy entanglement and the
harder to create it. Nevertheless, we see that even when $\Omega$ is
relatively large $\tau_G$ is still able to create considerable
amount of entanglement.

\section{Discussion and conclusion}
We have discussed the entangling and disentangled capacities of
nonlocal CPTP unital maps, i.e. maps that can be represented as
probabilistic mixtures of unitaries, and have calculated these
capacities in some particular cases for two qubits. Three classes of
unitaries were considered, namely the CNOT, DCNOT, and SWAP classes.

We have observed that the disentangling capacity was always larger
than corresponding entangling capacity, which contrasts with the
unitary case where the both capacities are equal for qubits and for
higher dimensions disentangling capacity cannot be greater than
entangling capacity \cite{ent_dis_unequal}.

In the case of the CNOT-class our results were obtained via
straightforward generalization of the method for CNOT-class
unitaries. We argue that the (dis)entangling capacity is achieved
when a map acts on the optimal pure state from the same family as in
the unitary case. Both discrete and continuous mixtures were
analyzed. In the case of the DCNOT and SWAP-class direct numerical
optimization was performed. We have conjectured that dimensions of
the local ancillas are equal to the dimensions of the original
particles, i.e. the ancillas were taken to be qubits.

A number of open question can be addressed in a future research.

 It will be interesting
and useful to prove (or disprove) the general conjecture that the
sizes of local ancillas can be taken equal to the sizes of original
particles.

Here we have calculated single-shot capacities. In the case of
unitaries it had been shown that in the asymptotic regime one cannot
do better \cite{BHLS_cap_Ham, asym_ent_cap}. It is important to
check whether this result holds in the non-unitary case.

In the case of DCNOT and SWAP-mixtures we performed maximization
over pure states only thereby obtaining lover bounds on
$E^{\uparrow}(\tau)$ and $E^{\downarrow}(\tau)$, but not their
actual values.

In our future research we will address the question of whether these
bounds are tight. It might be the case that optimal states for these
operations are mixed and, consequently, the capacities are higher
than we have calculated.

%{\it Idea: measure of non-locality of map - difference
%$E^{uparrow}-E^{downarrow}$}

\begin{acknowledgments}
This work was funded by the U.K. Engineering and Physical Sciences
Research Council, Grant No. EP/C528042/1, and supported by the
European Union through the Integrated Project QAP (IST-3-015848) and
SECOQC.
\end{acknowledgments}

\appendix
\section{}
We have used two-dimensional ancilla on each side. Consider a
general state of four qubits in the tensor-product of the
computational bases of the original particles $A$, $B$ and the
ancillary particles $A'$, $B'$
\begin{equation}
|\Psi\rangle_{AA'BB'}=\sum_{i,j,k,l}
c_{i,j,k,l}|i\rangle_A|j\rangle_{A'}|k\rangle_B|l\rangle_{B'}.
\end{equation}

There are $16$ terms in the above superposition with $16$ complex
amplitudes $c_{i,j,k,l}$, therefore $|\Psi\rangle$ can be
parameterized using $30$ real numbers (if we take into account the
global phase and normalization). We will parameterize it in the
following way \cite{parametrization}. First, to facilitate our
analysis it is easier to incorporate four indexes $i,j,k$ and $l$,
each of which runs from 0 to 1, into a single index, $x$, that runs
from 1 to 16. This can be done by using the formula
$x=8i+4j+2k+l+1$, which is essentially a formula for converting a
number from the Boolean representation to the decimal. Second, we
present amplitudes $c_x$ in the form
\begin{equation}
c_x=|c_x|e^{i\theta_x},
\end{equation}
where $\sum_{x=1}^{16}|c_x|^2=1$ and $\theta_1=0$. Third, we
introduce new parameters $\phi_x$ such that
\begin{equation}
c_x=\sin\phi_{x-1}\prod_{y=x}^{15}\cos\phi_y,
\end{equation}
where $\phi_0=\pi/2$. Thus the state $|\Psi\rangle$ is parameterized
by 30 angles. The advantage of this parametrization is that we
restrict their values only to the interval $[0,2\pi]$ that
simplifies numerics.

We proceed as follows. A program generates a vector of $30$ random
numbers in the interval $[0,2\pi]$. This is the initial state. We
then apply the non-local map and obtain a final state. We calculate
the value of the gain in entanglement $\Delta
S=S(\tau(\Psi)_{BB'})-S(\tau(\Psi)_{AA'BB'})-S(Tr_{AA'}|\Psi\rangle\!\langle\Psi|)$.
After that we vary the values of the random vector by a small amount
and repeat these calculations again, thereby obtaining a gradient of
the change in entanglement in that point. We move along the gradient
to obtain the next $|\Psi\rangle$, and the procedure is repeated.
Eventually, the program reaches the maximum where it stops.

%$\Delta S is convex overall as its E_re is convex while von N. E. E. is concave.
%As we maximize convex function over a subset of a convex set...

%%%%%%%%%%%%%%%%%%%%%%%
%global optimization?
%%%%%%%%%%%%%%%%%%%%%%%%


\begin{thebibliography}{20}
\bibitem{pure} It was explicitly shown \cite{lhl} that the optimization can be restricted to pure
states.
\bibitem{classification} Strictly speaking DCNOT-class and SWAP-class
should be unified under a single class if analyzed according to the
criteria of interconvertability under LOCC \cite{U->U}. In the
framework of (dis)entangling capacity it is useful to identify them
as separate classes, because their behavior differs qualitatively.
\bibitem{mixture_same_class} Here we assume that all unitaries in
Eq. (\ref{map}) belong to the same class, which is typical for real
situations where only parameters of the coupling are subject to
variations.
\bibitem{parametrization} This method is a partial adaptation of the method used
in Ref. \cite{VP:ent_m_pp} for calculation the relative entropy of
entanglement.
\end{thebibliography}
\end{document}